\documentclass[twoside,11pt]{article}

\usepackage{blindtext}

\usepackage[preprint]{jmlr2e}

\usepackage{todonotes}
\usepackage{listings}
\usepackage{xcolor}
\definecolor{codegreen}{rgb}{0,0.6,0}
\definecolor{codegray}{rgb}{0.5,0.5,0.5}
\definecolor{codepurple}{rgb}{0.58,0,0.82}
\definecolor{backcolour}{rgb}{0.95,0.95,0.92}
\lstdefinestyle{mystyle}{
    backgroundcolor=\color{backcolour},   
    commentstyle=\color{codegreen},
    keywordstyle=\color{magenta},
    numberstyle=\tiny\color{codegray},
    stringstyle=\color{codepurple},
    basicstyle=\ttfamily\footnotesize,
    breakatwhitespace=false,         
    breaklines=true,                 
    captionpos=b,                    
    keepspaces=true,                 
    numbers=left,                    
    numbersep=5pt,                  
    showspaces=false,                
    showstringspaces=false,
    showtabs=false,                  
    tabsize=2
}
\usepackage{lastpage}
\jmlrheading{26}{2025}{1-\pageref{LastPage}}{1/21; Revised 5/22}{9/22}{21-0000}{Xin Guo, Anran Hu, Matteo Santamaria, Mahan Tajrobehkar and Junzi Zhang}

\ShortHeadings{\texttt{MFGLib}: A Library for Mean-Field Games}{Guo, Hu, Santamaria, Tajrobehkar and Zhang}
\firstpageno{1}

\begin{document}

\title{\texttt{MFGLib}: A Library for Mean-Field Games}

\author{\name Xin Guo
        \thanks{Corresponding authors.}
        \thanks{All authors contributed equally to this work and are listed in alphabetical order.}
        \email xinguo@berkeley.edu \\
       \addr 
       University of California, Berkeley, IEOR\\
       \addr
       Amazon.com (Amazon Scholar)
       \AND
       \name Anran Hu\footnotemark[1] \footnotemark[2] \email ah4277@columbia.edu \\
       \addr Columbia University, IEOR
       \AND
       \name Matteo Santamaria\footnotemark[2] \email matteosantamaria@berkeley.edu\\  
       \name Mahan Tajrobehkar\footnotemark[2] \email mahan\_tajrobehkar@berkeley.edu\\
        \addr 
       University of California, Berkeley, IEOR
       \AND
       \name Junzi Zhang\footnotemark[1] \footnotemark[2] 
       \email junzizmath@gmail.com\\
       \addr
       Citadel Securities (Work done prior to joining Citadel Securities)}

\editor{My editor}

\maketitle

\begin{abstract}
Mean-field games (MFGs) are limiting models to approximate $N$-player games, with a number of applications. Despite the ever-growing numerical literature on computation of MFGs, there is no library that allows researchers and practitioners to easily create and solve their own MFG problems. The purpose of this document is to introduce \texttt{MFGLib}, an open-source Python library for solving general MFGs with a user-friendly and customizable interface. It serves as a handy tool for creating and analyzing generic MFG environments, along with embedded auto-tuners for all implemented algorithms. The package is distributed under the MIT license and the source code and documentation can be found at \url{https://github.com/radar-research-lab/MFGLib/}.
\end{abstract}


\section{Introduction}
Large population games are ubiquitous in real-world problems. 
As the number of players in the game grows, however, the computational complexity grows exponentially and it becomes notoriously hard to solve such problems. Mean-field games, pioneered by the seminal work of  \citep{huang2006large} and  \citep{lasry2007mean}, are limiting models of large-population symmetric games in the infinite-agent regime, which models the interaction among players by the interaction between a representative player and the state-action population distribution of all other players. The most commonly adopted solution concept in the game theory literature is Nash equilibrium (NE), a policy from which no agent will unilaterally deviate. It has been shown that the NE  policy of the mean-field game is an $\epsilon$-NE of the  $N$-player game \citep{saldi2018markov}, with $\epsilon=O(\frac{1}{\sqrt{N}})$ \citep{huang2006large}. In practice, games with small $N$ on the order of tens can be well approximated by MFGs \citep{guo2019learning, kizilkale2019integral, cabannes2021solving}. 


Recently, the literature of MFGs has experienced an exponential growth  both in theory and in practice. In particular, there has been a surge of interests in the computation and learning of NEs in MFGs, with wide applications including bid recommendation \citep{guo2019learning}, high frequency trading \citep{lehalle2019mean}, product pricing  \citep{guo2023general}, dynamic routing \citep{cabannes2021solving}, and animal behavior simulation \citep{perrin2021mean}. 
Nevertheless, implementations of these environments and algorithms are currently by and large provided mainly for paper reproducibility or experimental/internal-use purposes. 

This manuscript introduces \texttt{MFGLib}, an open-source Python library dedicated to solving NEs for generic MFGs with a user-friendly and customizable interface, aiming at promoting both applications and research of MFGs. On one hand, it facilitates the creation and analysis of arbitrary user-defined MFG environments with minimal prior knowledge of MFGs. On the other hand, it serves as a modular and extensible code base 
for the community to easily prototype and implement new algorithms and environments of MFGs as well as their variants and generalizations.\footnote{A pre-release version of \texttt{MFGLib} has been internally used by Amazon Advertising for research and production and was reported to serve well for their purposes.}

\paragraph{Related libraries.}
Various libraries have been developed for $N$-player games, such as \texttt{QuantEcon} \citep{quantecon}, \texttt{Nashpy} \citep{nashpy} and \texttt{ilqgames} \citep{fridovich2020efficient}. In contrast, 
only very limited tools focus on MFGs and are mainly for experimental and internal use and hence not suitable for general users with their own customized environments and problems.

Among these very few existing MFG libraries, \texttt{OpenSpiel} \citep{LanctotEtAl2019OpenSpiel}, a collection of environments and algorithms for research in reinforcement learning and planning in games, is the closest one to \texttt{MFGLib}. \texttt{OpenSpiel} has dedicated a module to MFGs implementing several environments and algorithms. However, it lacks customizability and a user-friendly API. In fact, according to its documentation, their code is still experimental and is only recommended for internal use. Other MFG libraries such as \texttt{gmfg-learning} \citep{gmfg-learning} and \texttt{entropic-mfg} \citep{entropic-mfg} 
are mainly developed to support the experimental results of a particular paper and are not suitable for general MFG experiments.

\section{Brief overview of \texttt{MFGLib}}
\texttt{MFGLib} is a handy tool for solving the ($\epsilon$)-NE of general user-defined MFG environments.\footnote{The current library is focused on discrete-time MFGs with a finite horizon and finite state and action spaces. See \citep{guo2024mf} for a formal exposition of the topic.} 
In what follows, we briefly discuss the design and features of this library. See the online documentation (\url{https://mfglib.readthedocs.io/en/latest}) for additional details.

\subsection{Library design}
\texttt{MFGLib} consists of two core modules: environments and algorithms, which can be developed and extended entirely independently of one another. Target environments and algorithms can be separately imported and instantiated from their corresponding modules. Users may implement their own environments and algorithms, or choose from among many built-in options. The algorithms can then be deployed to solve the environment instance.
The user-facing API of \texttt{MFGLib} is designed such that each aforementioned step can be performed using simple codes. 


\paragraph{Environments.} Users define new environments with the syntax:
\begin{lstlisting}[language=Python]
from mfglib.env import Environment
user_env = Environment(T=T, S=S, A=A, mu0=mu0, r_max=r_max, 
                       reward_fn=reward_fn, 
                       tranistion_fn=tranistion_fn)
\end{lstlisting}
Here \texttt{T,S,A,mu0,r\_max} are the time horizon, state space size, action space size, initial state distribution and max reward of the MFG, respectively. The rewards \texttt{reward\_fn} and transitions \texttt{transition\_fn} are the mappings from time (\texttt{t}) and population distribution (\texttt{L\_t}) to rewards and transitions tensors, which are callables\footnote{We allow both function and class callables here. But for simplicity, we focus on function implementations of transitions and rewards here.} 
with the following signatures: 
\begin{lstlisting}[language=Python,basicstyle=\fontsize{8}{9}\selectfont\ttfamily]
def reward_fn(env: Environment, t: int, L_t: torch.Tensor) -> torch.Tensor: ...
def transition_fn(env: Environment, t: int, L_t: torch.Tensor) -> torch.Tensor: ...
\end{lstlisting}


In addition, the library comes pre-loaded with ten environments \citep{cui2021approximately,perrin2020fictitious,perolat2021scaling,guo2019learning,guo2024mf} 
that can be directly instantiated as class methods of the \texttt{Environment} class: \texttt{Environment.left\_right()}. These pre-loaded environments accept optional environment-dependent arguments to give users fine-grained control over the time horizon, state space, rewards, etc. Many of the ten environments are commonly used test cases in numerical experiments, but \texttt{MFGLib} is the first project to provide them all in a single codebase. \texttt{MFGLib} therefore provides an important testbed for researchers to evaluate new algorithms on a common set of environments.


\paragraph{Solvers and evaluation.} Once an environment \texttt{env} is created, one can then instantiate a solver from the options provided by \texttt{MFGLib} with the following syntax:
\begin{lstlisting}[language=Python]
from mfglib.alg import MFOMO
# Can also use OccupationMeasureInclusion, OnlineMirrorDescent, PriorDescent or FictitiousPlay
alg = MFOMO(**kwargs) 
policies, expls, runtimes = alg.solve(env)
\end{lstlisting}
Here the outputs store the policy iterates, exploitability scores\footnote{The exploitability scores evaluate the closeness of the output policies to  NE solutions.} and cumulative runtimes over iterations. A formatted log of the iteration process is also optionally printed out in real-time to help users monitor the performance.






\paragraph{Auto-tuning.} Beyond what is shown above, \texttt{MFGLib} also provides an auto-tuning tool to automatically select the best algorithm hyperparameters if performance under the default settings is unsatisfactory. The auto-tuner is invoked with the following syntax, 
where \texttt{env\_suite} is a list of environment instances to be tuned on:
\begin{lstlisting}[language=Python]
from mfglib.tuning import GeometricMean
tune_result = mfomo.tune(metric=GeometricMean(), envs=env_suite)
mfomo_tuned = MFOMO.from_study(tune_result)
policies, expls, runtime = mfomo_tuned.solve(env)
\end{lstlisting}
See the next subsection for more details on the available arguments of the \texttt{tune} method. 

\subsection{Key features}
In this section we highlight the key features of \texttt{MFGLib}. 

\paragraph{High-dimensional representation of state and action spaces.} We use \texttt{PyTorch} tensors to represent policies, mean-fields, rewards, etc. Whenever possible, we opt to keep the state and action spaces in their original form without performing any transformation.  
For high dimensional spaces, instead of flattening them and representing them using one dimensional spaces, we keep the original spaces. This treatment yields higher interpretability and provides more flexible and simpler interactions with users. 

\paragraph{Implemented algorithms.} \texttt{MFGLib} implements five state-of-the-art MFG algorithms including (Damped) Fictitious Play \citep{perrin2020fictitious,perolat2021scaling}, Online Mirror Descent \citep{perolat2021scaling}, Prior Descent \citep{cui2021approximately}, Mean-Field Occupation Measure Optimization (MFOMO) \citep{guo2024mf}, and Mean-Field Occupation Measure Inclusion (MFOMI) \citep{hu2024mfomlonlinemeanfieldreinforcement}. We remark that the implemented algorithms include many other existing algorithms as special cases, such as fixed point iteration and GMF-V algorithms \citep{guo2019learning}. 


\paragraph{Embedded tuner.} Every implemented algorithm requires at least one hyperparameter. 
To simplify the tuning process, \texttt{MFGLib} endows its algorithms with a built-in tuner, which can be used to tune the hyperparameters on one single environment instance or across several instances (an environment suite). The tuner can also be seeded with multiple policy initializations. The tuners are based on Optuna \citep{akiba2019optuna}, an open source hyperparameter optimization framework used to automate hyperparameter search. \texttt{MFGLib} provides several convenient Optuna-compatible objectives. For instance, the code 
\begin{lstlisting}[language=Python]
OnlineMirrorDescent().tune(
    envs=[env_1, env_2, env_3],
    pi0s=[pi0_1, pi0_2, pi0_3],
    metric=GeometricMean(shift=1),
)
\end{lstlisting}
minimizes the shifted geometric mean of the exploitability scores across multiple environment/policy initializations for Online Mirror Descent. Users can additionally implement their own metrics with little difficulty. The \lstinline{tune} method provides the users with several additional optional arguments to adjust the tuning process, including max iterations (\lstinline{max_iter}), the absolute and relative tolerances for early stopping (\lstinline{atol, rtol}), the number of trials (\lstinline{n_trials}) and the maximum run-time of the whole tuning process (\lstinline{timeout}).



\paragraph{Code quality and accessibility.} \texttt{MFGLib} adheres to the highest standards of code quality. The library is regularly subjected to unit testing and static analysis through a continuous integration (CI) system. We employ the latest tools such as \texttt{black} and \texttt{ruff} to ensure that the source code remains clean and readable. We also strictly type-check the library with \texttt{mypy} to eliminate an entire class of type-safety bugs.

\texttt{MFGLib} welcomes outside contributors and is easily accessible to anyone familiar with Python, beginners and experts alike. The folder structure is simple and easy to navigate. Since \texttt{MFGLib} is a pure Python implementation, there is no complicated build process to speak of. We hope the structural simplicity encourages users to engage with the library and submit new algorithms and environments.

\newpage
\acks{The authors would like to thank Sareh Nabi, Rabih Salhab, and Lihong Li of Amazon,  Xiaoyang Liu of Columbia University, and Zhaoran Wang of Northwestern
University for their feedback on early versions of \texttt{MFGLib}. }



\bibliography{mfglib}

\end{document}